\documentclass[12pt]{article}

\usepackage[latin1]{inputenc}
\usepackage{amsmath,cite}
\usepackage{amsfonts}
\usepackage{amssymb,MnSymbol,slashed}
\usepackage{amsxtra}
\usepackage[margin=3cm]{geometry}
\usepackage{color}
\usepackage{hyperref}
\hypersetup{linktocpage=true}
\usepackage{graphicx}
\usepackage{placeins}
\usepackage{upgreek} 
\usepackage{mathrsfs}
\usepackage{accents}
\usepackage{multirow} 
\usepackage{tikz}

\numberwithin{equation}{section}
\newcommand{\beq}{\begin{equation}}
\newcommand{\eeq}{\end{equation}}
\def\be {\begin{equation}}
\def\ee {\end{equation}}
\def\bs#1\es{\begin{split}#1\end{split}}
\def\ba#1\ea{\begin{align}#1\end{align}}
\def\baed#1\eaed{\begin{aligned}#1\end{aligned}}
\def\bged#1\eged{\begin{gathered}#1\end{gathered}}
\def\bea{\begin{eqnarray}}
\def\eea{\end{eqnarray}}

\def\nn{\nonumber}

\def\a{\alpha}

\def\d{\delta}
\def\D{\Delta}
\def\e{\epsilon}

\def\F{\Phi}

\def\G{\Gamma}
\def\h{\eta}
\def\k{\kappa}

\def\m{\mu}
\def\n{\nu}
\def\o{\omega}
\def\O{\Omega}

\def\r{\rho}

\def\bls{\bigg [}
\def\brs{\bigg ]}

\newcommand{\tb}[1]{ \textrm{\tiny{{({#1})}}} {} }

\newcommand{\cK}{\mathcal{K}}
\newcommand{\cW}{\mathcal{W}}
\newcommand{\cG}{\mathcal{G}}

\newcommand{\cR}{\mathcal{R}}

\def\cA{{{\mathcal A}}}
\def\cO{{{\mathcal O}}}
\def\cM{\mathcal{M}} 
\def\cN{\mathcal{N}}
\def\cV{\mathcal{V}}

\def\cZ{\mathcal{Z}}

\newcommand{\sm}[1]{\text{{\tiny{$#1$}}}}

\def\Tr{\text{Tr}}

\def\pa{\partial}
\def\na{\nabla}
\def\fr{\frac}
\def\tfr{\tfrac}

\def\we{\wedge}
\def\ra{\rightarrow}

\def\lra{\leftrightarrow}

\def\tbzero{{\text{\tiny{(0)}}}}
\def\tbone{{\text{\tiny{(1)}}}}
\def\tbtwo{{\text{\tiny{(2)}}}}
\def\eppr{\alpha}

\newcommand{\wh}[1]{ {\hat{#1}}{} }
\newcommand{\til}[1]{ {\tilde{#1}} }

\let\foo\bar 
\renewcommand{\bar}[1]{ {\foo{  #1} }{} }

\newlength{\dhatheight}

\begin{document}
\baselineskip=16pt
\setlength{\parskip}{6pt}

\begin{titlepage}
\begin{flushright}
\parbox[t]{1.4in}{
\flushright  MPP-2014-574}
\end{flushright}

\begin{center}

\vspace*{1.5cm}

{\Large \bf The effective action of warped M-theory reductions }\\[.1cm]
{\Large\bf with higher derivative terms}\\[.1cm]
{\large \bf -- Part I --}

\vskip 1.5cm

\renewcommand{\thefootnote}{}

\begin{center}
 \normalsize 
 Thomas W. Grimm, Tom G. Pugh, Matthias Weissenbacher \footnote{{ grimm, mweisse, pught @ mpp.mpg.de}}
\end{center}
\vskip 0.5cm
Max Planck Institute for Physics,\\
F\"ohringer Ring 6, 80805 Munich, Germany
\end{center}

\vskip 1.5cm
\renewcommand{\thefootnote}{\arabic{footnote}}

\begin{center} {\bf ABSTRACT } \end{center}
M-theory accessed via eleven-dimensional supergravity admits globally consistent 
warped solutions with eight-dimensional compact spaces if background fluxes and 
higher derivative terms are considered. The internal background is conformally K\"ahler 
with vanishing first Chern class. We perturb these solutions 
including a finite number of K\"ahler deformations of the metric and 
vector deformations of the M-theory three-form. Special emphasis is 
given to the field-dependence of the warp-factor and the higher-derivative
terms. We show that the three-dimensional two-derivative effective action 
takes a surprisingly simple form in terms of a single higher-curvature
building block due to numerous non-trivial cancellations. Both the ansatz and the effective action 
admit a moduli dependent scaling symmetry of the internal metric. Furthermore, we 
find that the required departure from Ricci-flatness and harmonicity of the zero-mode eigenforms 
does not alter the effective theory.

\end{titlepage}

\newpage
\noindent\rule{\textwidth}{.1pt}		
\tableofcontents
\vspace{20pt}
\noindent\rule{\textwidth}{.1pt}

\setcounter{page}{1}
\setlength{\parskip}{9pt} 

\newpage
\section{Introduction and discussion}

Dimensional reductions of M-theory on compact eight-dimensional manifolds result in
three-dimensional effective 
theories with various amounts of supersymmetry. These reductions are both of conceptual 
as well as phenomenological interest.
A phenomenological investigation 
might be carried out when applying the M-theory to F-theory limit in 
order to lift the three-dimensional theories to four space-time dimensions for a certain class 
of eight-dimensional manifolds \cite{Vafa:1996xn}. From a phenomenological 
point of view, compactifications in which the 
effective theory preserves only small amounts of supersymmetry
are of particular interest. For example, compactifications of M-theory and F-theory preserving
four supercharges allow for background fluxes that can induce a  
four-dimensional chiral spectrum.

 A famous class of warped solutions 
with background fluxes was argued to exist in \cite{Becker:1996gj}. Global consistency, however, 
requires that, in a compact scenario with background fluxes, higher-derivative terms
in the eleven-dimensional action must also be included. It was subsequently shown  
that there are indeed solutions that solve the higher-derivative field equations \cite{Becker:2001pm}.
More precisely, one finds that the internal background is a conformally K\"ahler manifold 
with vanishing first Chern class, but a metric that is non-Ricci-flat even when allowing for a conformal
rescaling including the warp factor. This deviation is due to the possible non-harmonicity of
the third Chern-form in the leading order Ricci-flat metric \cite{Grimm:2014xva}. While a complete  
check of supersymmetry is still missing, it was shown in \cite{Grimm:2014xva} that 
a modification of the eleven-dimensional gravitino variations with higher curvature
terms based on \cite{Lu:2003ze,Lu:2004ng} vanishes on the warped background solutions. 
It was furthermore argued, that the warped background admits a
globally defined real two-form $J'$ and complex 
four-form $\Omega'$. Separating the warp-factor, the Killing spinor equations translate into 
first order differential constraints on these forms, with only
$d \Omega' =\overline{\cW}_5 \wedge \Omega'$ non-vanishing
for an exact one-form $\overline{\cW}_5$.\footnote{At the two-derivative level 
eleven-dimensional supergravity on $SU(4)$ structure manifolds
has recently been studied in \cite{Prins:2013wza}.The fact that the metric is no longer Ricci 
flat when higher derivative couplings and $\alpha'$-corrections 
are taken into account is a classical result for Calabi-Yau manifolds without background fluxes 
in string theory \cite{Nemeschansky:1986yx} and has been recently investigated for 
$Spin(7)$ and $G_2$ compactifications \cite{Becker:2014rea}.}

In this paper we study the three-dimensional effective action 
arising when perturbing the solutions considered in \cite{Becker:2001pm,Grimm:2014xva} 
by a finite number of K\"ahler deformations of the metric and 
vector deformations of the M-theory three-form.
More precisely, our starting point is the bosonic part of the eleven-dimensional supergravity action of 
\cite{Cremmer:1978km} corrected by the terms fourth order in the Riemann
curvature known since the works \cite{Duff:1995wd,Green:1997di,Green:1997as,Kiritsis:1997em,Russo:1997mk,Antoniadis:1997eg,Tseytlin:2000sf}, 
and the higher-derivative terms quadratic in the M-theory three-form found 
in \cite{Liu:2013dna}. Let us stress that there are important terms 
of the structure $(\hat \nabla\hat G)^2 \hat R^3$, where $\hat G$ is the M-theory four-form field-strength and 
$\hat R$ is the Riemann curvature tensor, that have not been fully determined.
They were argued to be given by a number of building blocks of index contractions \cite{PaperPartII} with
4-point amplitudes only determining part of the numerical prefactors. Remarkably, 
most of these unknown coefficients actually do not effect our computation and
we are able to suggest a fixation of the unknown coefficients up to one 
constant. This last constant might then be fixed by supersymmetry \cite{PaperPartII}. 
Clearly, the complete form of the $(\hat \nabla\hat G)^2 \hat R^3$ terms 
could also be determined by considering amplitudes with 5 and more external legs.

Given the eleven-dimensional action with higher-derivative terms 
we systematically construct the perturbed background 
order by order in a scale parameter $\alpha \propto \ell_{M}^{3}$, where $\ell_M$ is the 
eleven-dimensional Planck length. At zeroth order in $\alpha$ the background is 
simply a direct product of a Calabi-Yau fourfold without background fluxes and 
preserves four supercharges. At higher order in $\alpha$ the fluxes and higher 
curvature terms need to be included. The metric ansatz is modified 
and accordingly the mode expansion for K\"ahler structure perturbations 
of the metric and vector perturbations of the M-theory three-form is described in terms of forms non-harmonic in the zeroth order Calabi-Yau metric. We carefully 
keep track of all such modifications, but show that most of these modifications
eventually cancel in the final three-dimensional effective action. In fact, inserting 
the ansatz into the higher-derivative action, we find that the kinetic terms 
for the deformations and vectors in the three-dimensional effective theory 
can be expressed using a single higher-curvature building 
block $Z_{m\bar m n \bar n} = \frac{1}{4!} (\epsilon_{8} \epsilon_{8} R^\tbzero{}^3)_{m\bar m n \bar n}$, where 
$ R^\tbzero$ is the internal Riemann tensor in the zeroth order Calabi-Yau metric, 
see \eqref{def-Z} for the precise form of $Z$. Let us note that $Z_{m\bar m n \bar n}$
has the same symmetries as the Riemann tensor. It contracts with $R^{\bar m m \bar n n}$
to the Hodge-dual of the fourth Chern-form, and contracting any of the index pairs with the metric 
one finds expressions in terms of the third Chern-form. The equivalent 
quantity on a Calabi-Yau threefold was found to be important in \cite{Katmadas:2013mma}. 
It would be interesting to examine if $Z_{m\bar m n \bar n}$
plays a special role in describing the topology of the compact eightfold.

In addition to the complications arising from reducing 
higher-derivative terms in the action, a proper treatment 
of the warp-factor turns out to be crucial. Warped compactifications of 
M-theory and Type IIB  have been considered previously in \cite{Dasgupta:1999ss,Giddings:2001yu,Giddings:2005ff,Burgess:2006mn,Shiu:2008ry,Douglas:2008jx,Martucci:2009sf,Underwood:2010pm,Frey:2013bha,Martucci:2014ska}, and were argued to 
be crucial in a complete understanding of the M-theory to 
F-theory limit for minimally supersymmetric setups \cite{Grimm:2012rg}.
In this work we perform the crucial generalization 
to include the higher-derivative terms, since warped compactifications 
with fluxes are inconsistent without these contributions. 
It turns out, that in this general case the modifications of the 
warp-factor to the lower-dimensional effective theory 
are significantly more involved then the ones discussed previously in the literature. 
Nevertheless we will be able to show that the effective theory permits a
non-trivial scaling symmetry induced by rescaling the warp-factor by a field-dependent function. 
In a subsequent paper \cite{PaperPartII} we will argue that the 
three-dimensional action carries the properties of 
a $\mathcal{N} = 2$ supergravity theory and extend 
the results of \cite{Grimm:2013gma,Grimm:2013bha,Junghans:2014zla}.

The paper is organized as follows. In section \ref{action+solution} we 
review the eleven-dimensional effective action of M-theory including 
higher-derivative terms. We then introduce the considered warped 
solutions that admit an eight-dimensional compact internal manifold
and background fluxes and comment on the supersymmetry conditions. 
The considered perturbations of the background solutions 
are introduced in section \ref{Perturbations_back} and consist of vector modes 
of the M-theory three-form and K\"ahler structure deformations. We 
also discuss the field-dependence of the warp-factor. The
dimensional reduction yielding a three-dimensional effective 
action is carried out in section \ref{3Daction}, where we present the 
results for the kinetic terms and Chern-Simons terms. 
A summary of our conventions and a number of useful identities are 
supplemented in appendix \ref{Conventions}. More details on 
the dimensional reduction of the higher derivative terms 
can be found in appendix \ref{ReductionResults}.

\section{Eleven-dimensional action and compactifying solutions} \label{action+solution}

In this section we introduce the eleven-dimensional action including 
the known higher-derivative terms that will then be used in the dimensional 
reduction. The individual terms are discussed in subsection \ref{11Daction},
with details and conventions supplemented in appendix \ref{Conventions}. The eleven-dimensional
theory admits a warped solution with a compact eight-dimensional space and background
fluxes as we recall in subsection \ref{11Dsolution}.

\subsection{The eleven-dimensional action with higher-derivative terms} \label{11Daction}

Our starting point will be the eleven-dimensional supergravity action that arises as the 
low energy limit of M-theory. At the two-derivative level the action is the 
long-known $\cN=1$ supergravity action first worked out in \cite{Cremmer:1978km}.
Recall that the dynamical fields of this supergravity theory 
arrange in an $\cN=1$ gravity multiplet, with bosonic fields being the eleven-dimensional metric 
$\hat g_{NM}$ and a three-form $\hat C_{MNP}$ with field strength $\hat G_{QMNP} = \partial_{[Q}\hat C_{MNP]}$.
In the following we will indicate eleven-dimensional quantities with a hat.
The action for these bosonic fields is given by
\ba
S^\tbzero &=  \fr{1}{2 \k_{11}^2}  \int \Big[
\wh R \wh * 1 - \fr12 \wh G \we \wh * \wh G - \fr16 \wh C \we \wh G \we \wh G \Big] \ ,
\ea
where $\hat R$ is the Ricci scalar evaluated with conventions introduced in appendix \ref{Conventions}.

In order to find globally consistent solutions with internal background fluxes for $\wh G$ 
one has to include higher-derivative corrections to the theory as we recall below. 
Terms that are up to eighth order in derivatives and are quadratic in 
$\wh G$ will be crucial in this discussion. To systematically display the results we introduce the 
dimensionful parameter 
\beq
  \eppr{}^2 = \fr{( 4 \pi \k_{11}^2)^\fr23}{(2 \pi)^4 3^2 2^{13}} \, . 
\eeq
These bosonic terms have been worked out in \cite{Duff:1995wd,Green:1997di,Green:1997as,Kiritsis:1997em,Russo:1997mk,Antoniadis:1997eg,Tseytlin:2000sf,Liu:2013dna},
such that the action takes the form 
\ba \label{complete_11Daction}
S & = S^\tbzero + \eppr^2 S^\tbtwo_{\wh R^4}  +   \eppr^2 S^\tbtwo_{\wh G^2 \wh R^3 } +  \eppr^2 S^\tbtwo_{(\wh \na \wh G)^2  \wh R^2} + \cO (  \wh G^3 \eppr^2 ) + \cO( \eppr^3)  \, , 
\ea
with eight-derivative terms given by 
\ba
S^\tbtwo_{\wh R^4} &=  \fr{1}{2 \k_{11}^2}  \int \Big[  
(\wh t_8 \wh t_8 - \fr1{24} \wh \e_{11} \wh \e_{11} ) \wh R^4 \wh * 1 
- 3^2 2^{13}  \wh C \we \wh X_8 \Big] \ , \label{M-TheoryAction_1} \\
S^\tbtwo_{\wh G^2 \wh R^3} & = \fr{1}{2 \k_{11}^2}  \int \Big[ -   (\wh t_8 \wh t_8 + \fr1{96} \wh \e_{11} \wh \e_{11} ) \wh G^2 \wh R^3 \wh * 1 \Big]\ , \label{M-TheoryAction_2} \\
S^\tbtwo_{(\wh \na \wh G)^2 \wh R^2} & = \fr{1}{2 \k_{11}^2}  \int \wh s_{18} (\wh \na \wh G)^2 \wh R^2  \wh * 1 \ .
\label{M-TheoryAction_3}
\ea
The terms at higher order in $\wh G$ and $\eppr$ will not be needed in what follows 
as their contribution is higher order in $\eppr$ when evaluated on the ansatz we will make. 

Let us now discuss the various couplings in \eqref{M-TheoryAction_1}-\eqref{M-TheoryAction_3} in more detail.
In \eqref{M-TheoryAction_1} we make the definitions 
\ba
 \wh X_8 = \fr{1}{192} \Big( \Tr \wh \cR^4 - \fr14 (\Tr \wh \cR^2)^2 \Big) \ ,
\ea
where $\wh \cR$ is the eleven-dimensional curvature two-from $\wh \cR^{M}_{\ \ N} = \frac12 \wh R^{M}_{\ \ NPQ} dx^P \wedge dx^Q$,
and 
\ba
 \wh \e_{11} \wh \e_{11} \wh R^4 &=  \epsilon^{R_1 R_2 R_3  M_1\ldots M_{8} }  \epsilon_{R_1 R_2 R_3 N_1 \ldots N_{8}} \wh R^{N_1 N_2}{}_{M_1 M_2} \wh R^{N_3 N_4}{}_{M_3 M_4}  \wh R^{N_5 N_6}{}_{M_5 M_6} \wh R^{N_7 N_8}{}_{M_7 M_8} \ , \nn \\[.1cm]
 \wh t_8 \wh t_8 \wh R^4 & = \wh t_{8}^{  M_1 \dots M_8} \wh t_{8 \, N_1  \dots N_8}     \wh R^{N_1 N_2}{}_{M_1 M_2} \wh R^{N_3 N_4}{}_{M_3 M_4}  \wh R^{N_5 N_6}{}_{M_5 M_6} \wh R^{N_7 N_8}{}_{M_7 M_8} \ , 
\ea
where $\e_{11}$ is the eleven-dimensional totally anti-symmetric epsilon tensor and 
$t_8$ is given explicitly in \eqref{def-t8} in appendix \ref{Conventions}.
Using $\e_{11}$ and $t_8$ the explicit form for the terms in \eqref{M-TheoryAction_2}
is given by 
\ba
\wh  \epsilon_{11} \wh \epsilon_{11} \wh G^2 \wh R^3 &= \wh \epsilon^{R M_1 \ldots M_{10} } \wh \epsilon_{R N_1 \ldots N_{10}}  \wh G^{N_1 N_2}{}_{M_1 M_2} \wh G^{N_3 N_4}{}_{M_3 M_4}    \wh R^{N_5 N_6}{}_{M_5 M_6} \wh R^{N_7 N_8}{}_{M_7 M_8} \wh R^{N_9 N_{10}}{}_{M_9 M_{10}} \,, \nn \\[.1cm]
 \label{eq:ttGR}
\wh  t_8 \wh t_8 \wh G^2 \wh R^3  & = \wh t_8^{ M_1  \dots M_8}  \wh t_{8}{}_{  N_1 \dots N_8}  \wh G^{N_1}{}_{M_1}{}_{R_1 R_2} \wh G^{N_2}{}_{M_2}{}^{R_1 R_2}  \wh R^{N_3 N_4}{}_{M_3 M_4}  \wh R^{N_5 N_6}{}_{M_5 M_6} \wh R^{N_7 N_8}{}_{M_7 M_8} \, . 
\ea

Finally, we need to introduce the tensor $\wh s_{18}^{N_1 \ldots N_{18}} $ appearing in \eqref{M-TheoryAction_3}.
Unfortunately, the precise form of $\wh s_{18}$ is not known. However, one can fix significant parts of it 
following \cite{Peeters:2005tb}. In order to express these parts we use the basis $B_i,\ i=1,...,24$ of \cite{Peeters:2005tb}, 
that labels all unrelated index contractions in $\wh s_{18} (\wh \na \wh G)^2 \wh R^2 $.
The $B_i$ are explicitly given in \eqref{def-Bi}. 
The result can then be expressed in terms of a 4-point amplitude contribution $\cA$ 
and a linear combination of six contributions $\cZ_i$ which do not affect the 4-point amplitude as
\ba \label{s18term_exp}
 \wh s_{18} (\wh \na \wh G)^2 \wh R^2 = \wh s_{18}^{N_1 \ldots N_{18}} \wh R_{N_1\ldots N_{4}} \wh R_{N_5\ldots N_{8}} \wh \na_{N_9} \wh G_{N_{10} \ldots N_{13} } \wh \na_{N_{14}} \wh G_{N_{15} \ldots N_{18}} = \cA + \sum_n a_n \cZ_n \, . 
\ea
The combinations $\cA$ and $\cZ_n$ are then given in terms of the basis elements as
\ba
\cA &= -24B_{5} -48B_{8} -24B_{10} -6B_{12} -12B_{13} +12B_{14}
 +8B_{16} -4 B_{20} + B_{22} + 4B_{23} + B_{24}  \,, \nn \\
\cZ_1 &= 48B_1 + 48B_2 - 48B_3 + 36B_4 + 96B_6 + 48B_7 - 48B_8 +
96B_{10} \nn \\
&\quad\quad + 12B_{12} +24B_{13} -12B_{14} + 8B_{15} + 8B_{16} - 16B_{17} + 6B_{19} + 2B_{22} + B_{24}\,,\nn \\
\cZ_2 &=-48B_1 -48B_2 -24B_4 -24B_5 +48B_6 -48B_8 -24B_9
-72B_{10}  -24B_{13} +24B_{14} -B_{22} +4B_{23}\,,\nn \\
\cZ_3 &= 12B_1 + 12B_2 - 24B_3 + 9B_4 +48B_6 + 24B_7 - 24B_8
 + 24B_{10}
 \nn \\
& \quad \quad
+ 6B_{12} 
 + 6B_{13}  + 4B_{15} - 4B_{17} + 3B_{19} + 2B_{21}\,,\nn \\
\cZ_4 &= 12B_1 + 12B_2 - 12B_3 + 9B_4 +24B_6 + 12B_7 - 12B_8 + 24B_{10}
+ 3B_{12} + 6B_{13} + 4B_{15} - 4B_{17} + 2B_{20}\,,\nn \\
\cZ_5 &= 4B_{3} -8B_{6} -4B_7 + 4B_8 -B_{12} -2B_{14} + 4B_{18}\,,\nn \\
\cZ_6 &= B_4 + 2B_{11} \,.
\ea
We will show in this work that the terms 
$\cZ_3$ to $\cZ_6$ vanish both on the considered background solution 
and their perturbed cousins to the order in $\alpha$ we are considering.
In the next subsection we discuss the solutions in 
more detail.

\subsection{Compactifying warped solutions with background fluxes} \label{11Dsolution}

In the following we will review the warped solutions following \cite{Becker:2001pm,Grimm:2014xva}.
The starting point are the field equations derived from the action \eqref{complete_11Daction}.
These have a solution with an eleven-dimensional metric background 
\ba \label{MetricAnsatzeps2}
 d \wh s^2 &= e^{\eppr^2  \F^\tbtwo } ( e^{-2 \eppr^2 W^\tbtwo}   \h_{\m\n} dx^\m dx^\n  +  2 e^{  \eppr^2 W^\tbtwo}  g_{m \bar n} dy^m dy^\bar n )\ + \cO(\eppr^3) ,
\ea
where $\h_{\m\n} $ is the three-dimensional Minkowski metric and  
\ba 
g_{m \bar n} = g^\tbzero_{m \bar n} + \eppr^2 g^\tbtwo_{m \bar n} + \cO(\eppr^3)\, .
\label{gexp}
\ea
In the following we will denote the internal compact manifold by $Y_4$.
Here $\F^\tbtwo $ and $W^\tbtwo$ are scalar function on the internal space. $\F^\tbtwo$ represents 
an eleven-dimensional Weyl rescaling that will be given in terms of the internal space 
Riemann tensor below. $W^\tbtwo$ is known as the warp-factor and 
generally cannot be given explicitly, but rather is constraint by a differential equation \eqref{warpfactoreq}
known as the warp-factor equation.
In order to give the expansion \eqref{gexp} we note that 
at zeroth order in $\alpha$ the background is a direct product and 
$g^\tbzero_{m \bar n}$ is a Ricci flat metric. In fact, 
supersymmetry of the background at lowest order in $\eppr$  
demands that the metric $g^\tbzero_{m \bar n}$ must be that of a Calabi-Yau fourfold. 
We therefore can introduce complex indices, which here and in the following 
always refer to the zeroth order complex structure on the internal manifold. 
On a Calabi-Yau fourfold there exists a nowhere vanishing covariantly constant 
K\"ahler form $J^\tbzero$ and holomorphic $(4,0)$-form $\O^\tbzero$ satisfying
\ba
d J^\tbzero = d \O^\tbzero = 0 \, . 
\ea 
In what follows we will work in conventions in which the internal space indices are 
raised and lowered with the lowest order internal space metric $g^\tbzero_{m \bar n} $.

The background also includes a flux for the four-form given by
\ba
 \wh G_{m \bar n r \bar s} & = \eppr G_{m \bar n r \bar s}^\tbone + \cO(\eppr^3) \, , &
 \wh G_{m n r s} & = \eppr G_{m n r s}^\tbone + \cO(\eppr^3) \, , \nn 
\ea
\vspace{-1cm}
\ba \label{FluxAnsatz}
 \wh G_{\m\n\r m} &= \e_{\m \n \r} \pa_m e^{ -3 \eppr^2 W^\tbtwo} + \cO(\eppr^3)  \,  .
\ea
In order that the eleven-dimensional field equations are solved to order $\alpha^2$ by 
this background the flux $G^\tbone $ must be 
self-dual in the lowest-order metric $g^\tbzero_{m \bar n}$. This condition allows 
$(2,2)$ and $(4,0)+(0,4)$ components of the flux with respect to the lowest order complex structure.

The analysis of the higher derivative equations of motion fixes the value of the 
eleven-dimensional Weyl rescaling $\F^\tbtwo$ in terms of the lowest order metric $g^\tbzero_{m \bar n}$ as 
\ba
\F^\tbtwo &= - \tfr{512}{3} Z\, , & 
Z &= *^\tbzero ( J^\tbzero \we c_3^\tbzero )\ ,
\ea
where $c^\tbzero_3$ is the third Chern form built from $g^\tbzero_{m \bar n}$. As $c^\tbzero_3$ is a closed real six-form on a K\"ahler manifold we may write 
\ba
c_3^\tbzero &= H^\tbzero c_3^\tbzero + i \pa^\tbzero \bar \pa^\tbzero F \, , 
\ea
where $H^\tbzero$ indicates the projection to the harmonic part associated with the metric $g^\tbzero_{m \bar n }$. 
Using this decomposition we note that the scalar $Z$ is given by
\ba
Z &= *^\tbzero ( J^\tbzero \we H^\tbzero c_3^\tbzero) + \tfr{1}{4} \D^\tbzero *^\tbzero ( J^\tbzero \we J^\tbzero \we F) \, . 
\ea
The higher-derivative Einstein equations then fix the metric correction to be 
\ba
\label{g2Expression}
g_{m \bar n}^{\tbtwo} &= 768\pa_m^\tbzero \bar \pa_{\bar n}^\tbzero \til F\ , & 
 \til F &= *^\tbzero ( J^\tbzero \we J^\tbzero \we F) \ .
\ea
This implies that the metric $g_{m \bar n}$ introduced in \eqref{gexp} is 
still K\"ahler and that the internal part of the eleven-dimensional metric \eqref{MetricAnsatzeps2}
is conformally K\"ahler.
The field equations for the M-theory three-form $\hat C$ and the external space 
Einstein equations then constrain the warp-factor $W^\tbtwo$ to satisfy 
\ba \label{warpfactoreq}
  & d^\dagger d e^{ 3 \alpha^2 W^\tbtwo}  + \alpha^2 \fr12  G^\tbone \we G^{\tbone}
+ 3^2 2^{13} \alpha^2 X_{8} + \cO(\alpha^3) = 0\ . 
\ea
With these expressions one can demonstrate that all eleven-dimensional 
equations of motion are indeed satisfied \cite{Becker:2001pm,Grimm:2014xva}. 
For a compact $Y_4$ the warp-factor equation \eqref{warpfactoreq} implies the global consistency condition
\beq \label{tadpole}
   \frac{1}{3^2 2^{14}} \int_{Y_4} G^\tbone \we G^{\tbone} =  \frac{\chi(Y_4)}{24} \ ,
\eeq
where $\chi(Y_4) = - 4! \int_{Y_4} X_8 $ is the Euler number of $Y_4$.
Using self-duality of the fluxes $G^{\tbone}$ one thus realizes that in 
higher-derivative terms cannot be consistently ignored if one 
allows for a background flux. The somewhat unusual numerical 
factor in \eqref{tadpole} stems from our normalization of $G^{\tbone}$
with $\alpha$ and can be removed when moving to quantized 
fluxes $G^{\rm flux} = \frac{1}{3\, 2^6 \sqrt2 } G^\tbone$.

Let us close this section with a short discussion on supersymmetry. 
It should be stressed that the full supersymmetric completion of the 
action \eqref{complete_11Daction} is not known and neither have the supersymmetry variations
of the fermions been written down. In \cite{Grimm:2014xva} a proposal 
was made for the gravitino variations including order $\alpha^2$-terms based on \cite{Lu:2003ze,Lu:2004ng}.
It was shown to be compatible with the Einstein equations. At linear
order in $\alpha$ the supersymmetry variations were unchanged 
and the condition on the flux is the vanishing 
of the $(4,0)+(0,4)$-component of $G^\tbone$, i.e.
\beq \label{G40}
   G^\tbone_{mnrs} = 0\ ,
\eeq
and the primitivity condition 
\beq \label{primitivitycond}
 G^\tbone \we J^\tbzero = 0\ .
\eeq
It was also argued in \cite{Grimm:2014xva} that the presented solution
for the metric is compatible with the proposed Killing spinor equations at order $\alpha^2$. 
Since we will not bring the three-dimensional effective action into standard $\cN=2$ form,
the discussion of supersymmetry will not be crucial in this work. 

\section{Perturbations of the background} \label{Perturbations_back}

In subsection \ref{11Dsolution} we have reviewed a supersymmetric background with an
internal compact space that is conformally K\"ahler. We will now examine a set of deformations 
that preserve the K\"ahler condition but change the chosen K\"ahler structure. Our whole 
discussion will be carried out at fixed complex structure, i.e.~there are no complex structure 
deformations that will be switched on. In the following, the complex structure is chosen 
such that the supersymmetry condition \eqref{G40} on the flux is satisfied.
At lowest order in $\a$ the K\"ahler structure 
deformations are known to combine with vectors arising from the M-theory three-form $\wh C$
into three-dimensional $\cN=2$ multiplets, as discussed e.g.~in \cite{Haack:1999zv,Haack:2001jz}. 
We therefore need to study vectors arising from $\hat C$ taking into account 
higher $\alpha$-corrections in subsection~\ref{vectormod}.
The real scalars $v^i$ that correspond to the deformations of the K\"ahler structure 
will be introduced in subsection~\ref{Kahlermod}. In this latter subsection we will 
also study the variations of the warp-factor equation with respect to the K\"ahler structure
deformations. 

\subsection{Vector modes from the M-theory three-form} \label{vectormod}

Let us first examine the vector which arises in perturbations of the M-theory three-form 
$\wh C$. These correspond to a extra terms
in the expansion of $\wh G$ of the form 
\beq \label{deltaG}
   \d \wh G = F^i \we \o_i^\tb{v} \,, 
\eeq
where $F^i = d A^i$ and so provides the field strength for a three-dimensional vector $A^i$,
and $\o_i^\tb{v}$ are two-forms on the internal manifold. 
The tensor gauge symmetry of $\wh G$ translates to the $U(1)$ gauge symmetry of 
the $A^i$  in the three-dimensional effective theory. 

In order to make the meaning of \eqref{deltaG} precise, we need to specify the 
two-forms $\o_i^\tb{v}$. Therefore, as with the background fields studied in subsection \ref{11Dsolution},
we consider the expansion of $\o_i^\tb{v}$ to order $\alpha^2$ as
\ba
\o_i^\tb{v} = \o_i^\tbzero{}^\tb{v} + \eppr^2  \o_i^\tbtwo{}^\tb{v}\ .
\label{omexpansionv}
\ea
By making use of the Bianchi identity $d \wh G = 0$ in the absence of localized sources 
we see that $d \o_i^\tbzero{}^\tb{v} = d \o_i^\tbtwo{}^\tb{v} = 0$. 
The standard analysis of the lowest order reduction shows that only the harmonic 
part of $\o_i^\tbzero{}^\tb{v}$ contributes in the effective action and therefore we 
may pick $\o_i^\tbzero{}^\tb{v}$ to be harmonic. On a Calabi-Yau fourfold this implies 
that $\o_i^\tbzero{}^\tb{v}$ is a $(1,1)$-form and one has $i=1,\ldots,\text{dim}(H^{1,1}(Y_4))$,
where $H^{1,1}(Y_4)$ is the $(1,1)$-form cohomology of $Y_4$ whose  dimension 
is independent of the metric chosen on $Y_4$.

Let us next turn to $\o_i^\tbtwo{}^\tb{v}$. We first note that  
$\o_i^\tbzero{}^\tb{v}$ can be redefined to absorb the harmonic part of $\o_i^\tbtwo{}^\tb{v}$. 
This implies that $\o_i^\tbtwo{}^\tb{v}$ must be exact and as it is a real two-form 
on a K\"ahler manifold the $\pa \bar \pa$-lemma implies that it can be 
obtained by a $\pa^\tbzero \bar \pa^\tbzero$ of a scalar $ \r_i^\tb{v}$.
In other words, one can write 
\ba
\o_i^\tbzero{}^\tb{v} &= H^\tbzero \o_i^\tbzero{}^\tb{v} \,,  & 
\o_i^\tbtwo{}^\tb{v} &= \pa^\tbzero \bar \pa^\tbzero \r_i^\tb{v}\ .
\ea
The scalars $\r_i^\tb{v}$ parametrizes our ignorance in incorporating the 
higher-derivative corrections in the ansatz for the three-dimensional vector perturbations.  
Strictly speaking the indices $i$ on the $\r_i^\tb{v}$ and hence $\o_i^\tbtwo{}^\tb{v}$ 
and $\o_i^\tb{v}$ are not restricted to the range $1,\ldots, \text{dim}(H^{1,1}(Y_4))$
as before. However, as we will see in the explicit derivation of the effective 
action, all $\r_i^\tb{v}$ actually drop out of the final expression and 
therefore cannot yield additional dynamical fields. 
Interestingly, there is also a particular choice $\r_i^\tb{v}$ one could imagine,  
where $\o_i^\tb{v}$ is harmonic with respect to the full internal space metric \eqref{MetricAnsatzeps2}. 

\subsection{K\"ahler structure deformations and the warp-factor} \label{Kahlermod}

We now turn to the study of K\"ahler structure deformations
of the conformally K\"ahler metric in \eqref{MetricAnsatzeps2}. In 
order to do that, we introduce variations 
\beq
\d g_{m \bar n} = i \d v^i \o_{i \, m \bar n }^\tb{s}\ ,
\eeq
where $ g_{m \bar n}$ is the K\"ahler metric given in \eqref{gexp}. The $\d v^i$ correspond 
to scalars in the three-dimensional effective theory, while the $ \o_{i \, m \bar n }^\tb{s}$
is a set of two-forms on $Y_4$. Despite the misuse of notation, the field-range of the index $i$ is 
not yet restricted. The key point is to consider only $\o_{i \, m \bar n }^\tb{s}$
that preserve the K\"ahler condition. 
As before we can expand the forms $\o_i^\tb{s}$ in $\eppr$ as
\ba
\o_i^\tb{s} = \o_i^\tbzero{}^\tb{s} + \eppr^2  \o_i^\tbtwo{}^\tb{s}\ .
\label{omexpansions}
\ea
Preserving the K\"ahler condition requires that we impose $d \o_i^\tbzero = d \o_i^\tbtwo = 0$.
As before, we recall that at zeroth order in the parameter $\eppr$ 
the fluctuations $\d v^i$ are the well-known K\"ahler structure deformations
of the Calabi-Yau metric $g^\tbzero_{m \bar n} $ and 
the $\o_i^\tbzero{}^\tb{s}$ can be chosen to be harmonic $(1,1)$-forms with $i=1,\ldots, \text{dim}(H^{1,1}(Y_4))$. 
We may then make a redefinition to absorb the harmonic part 
of $\o_i^\tbtwo{}^\tb{s}$ so that $\o_i^\tbtwo{}^\tb{s} =\pa^\tbzero \bar \pa^\tbzero \r_i^\tb{s}$. We may then 
redefine the $\d v^i$ such that the lowest order harmonic $(1,1)$-forms match those used in the vector case 
\beq
  \o_i^\tbzero{}^\tb{s} = \o_i^\tbzero{}^\tb{v} = \o_i^\tbzero{}\ .
\eeq
Importantly the range of the index on the $\r_i^\tb{s}$ is once again a priori not restricted and 
there could be many more $\d v^i$ than harmonic forms. However, we will again see that 
all the $\r_i^\tb{s}$ as well as $\tilde F$ appearing in \eqref{g2Expression} do not appear in 
the three-dimensional effective action. 
This implies that one can equally consider deformations of the form 
\beq \label{g0shift}
 \d g^\tbzero_{m \bar n} = i \d v^i \o^{\tbzero}_{i \, m \bar n }\ ,
 \eeq
while making sure that all other quantities in the ansatz that are built from $g^\tbzero_{m \bar n} $ 
shift accordingly. 
It will be also convenient to define scalars $v^i$ containing the background value 
of $g^\tbzero_{m \bar n}$ by setting 
\beq \label{flucg}
  g^\tbzero_{m \bar n} +  \d g^\tbzero_{m \bar n} = i v^i \o^{\tbzero}_{i \, m \bar n }
\eeq

There are two main complications that arise when discussing the K\"ahler structure
deformations in a warped flux compactification. Firstly, they will in general 
not all be massless. Secondly, a change of K\"ahler structure will 
induce a shift in the warp-factor. 
The first of these points is seen at linear order in $\eppr$.
When the shift \eqref{g0shift} is made we see that the 
primitivity condition $G^\tbone \we J^\tbzero = 0$ given in \eqref{primitivitycond} is not 
preserved by the full set of fluctuations. This means that for constant $ \d v^i$ the field 
equations do not remain solved and so the 
full range of $ \d v^i$ no longer represent massless moduli of the background. Instead the 
set of massless $ \d v^i$ now becomes those that satisfy 
\beq \label{flucprimitive}
   \d v^i \o_i^\tbzero \we G^\tbone = 0\ . 
\eeq
These terms are responsible for the well 
known potential terms studied in the Calabi-Yau fourfold reductions with fluxes in \cite{Haack:1999zv,Haack:2001jz}. 
That this result for the potential is not effected by the higher-order corrections that result from 
higher-curvature terms is due to the fact that the supersymmetry conditions receive no linear modification in $\alpha$
and the potential is the square of this supersymmetry constraint. 

Let us now focus on the warp-factor. Going to second order in $\eppr$ 
we find that in addition to \eqref{FluxAnsatz} the fluctuations $\d v^i$ 
must also preserve the warp factor equation \eqref{warpfactoreq}. In order that this equation is preserved by the 
fluctuations we must now take the warp-factor to depend both on the internal space position 
and also the fields $\d v^i$ such that $W^\tbtwo = W^\tbtwo(y^m, v^i)$. 
When we perturb the background we will then find that the derivatives of $W^\tbtwo$ with respect to $v^i$,  
denoted by $\partial_i W^\tbtwo$, appear in these equations. We will only deduce the effective action for the 
fluctuations $\d v^i$ up to second order in $\d v^i$ and therefore it will suffice to consider $W^\tbtwo$ to be described by the truncated Taylor series
\ba
W^\tbtwo(y^m, v^i) = W^\tbtwo | + \pa_i W^\tbtwo | \d v^i + \fr12 \pa_i \pa_j W^\tbtwo | \d v^i \d v^j\ ,
\ea
where $W^\tbtwo |$ indicates the restriction of $W^\tbtwo$ to the point in moduli space where $ \d v^i = 0$.
Demanding that \eqref{warpfactoreq} is invariant up to second order in $\d v^i$ we find that
at first order in $\d v^i$ one has to impose
\ba \label{first_order_warp}
\na^\tbzero{}^m \na^\tbzero{}^{\bar n} (  g^\tbzero_{m \bar n } \pa_i W^\tbtwo \big|- i \o^\tbzero_{i m \bar n } W^\tbtwo | +i \o^\tbzero_{i}{}^r{}_r g_{m \bar n } W^\tbtwo |- i 2048 \o^\tbzero_i{}^{\bar s r} Z_{m \bar n r \bar s} ) = 0 \ ,
\ea
while at second order one constrains  
\ba \label{second_order_warp}
& \na^\tbzero{}^m \na^\tbzero{}^{\bar n} (   g^\tbzero_{m \bar n } \pa_i \pa_j W^\tbtwo \big|  - 2 i \o^\tbzero_{(i| m \bar n} \pa_{|j)} W^\tbtwo | -2 \o^\tbzero_{(i| m \bar s } \o_{|j)}^{\bar s}{}_{ \bar n } W^\tbtwo |
+ \o^\tbzero_{i}{}^r{}_r \o^\tbzero_{j}{}^s{}_s g^\tbzero_{m \bar n } W^\tbtwo |  \nn \\ 
& 
+ \o^\tbzero_{i}{}^r{}_s \o^\tbzero_{j}{}^s{}_r g^\tbzero_{m \bar n } W^\tbtwo | 
- 4096 \o^\tbzero_i{}^{\bar s r} \o^\tbzero_i{}^{\bar t}{}_{\bar t} Z_{m \bar n r \bar s} 
- 2048 \o^\tbzero_i{}^{\bar s t} \o^\tbzero_{i t}{}^{r} Z_{m \bar n r \bar s} + 6114 Y_{i j m \bar n} 
) = 0 \ .
\ea
In these variational constraints we have defined
\ba \label{def-Z}
Z_{m \bar m n \bar n} = \fr{1}{4!} \e^\tbzero_{ m \bar m m_1 \bar m_1 m_2 \bar m_2 m_3 \bar m_3}  \e^\tbzero_{ n \bar n n_1 \bar n_1 n_2 \bar n_2 n_3 \bar n_3} R^\tbzero{}^{\bar m_1 m_1 \bar n_1 n_1} R^\tbzero{}^{\bar m_2 m_2 \bar n_2 n_2} R^\tbzero{}^{\bar m_3 m_3 \bar n_3 n_3}\ ,
\ea
and 
\ba
Y_{i j m \bar n} = \fr{1}{4!} \e^\tbzero_{ m \bar m m_1 \bar m_1 m_2 \bar m_2 m_3 \bar m_3}  \e^\tbzero_{ n \bar n n_1 \bar n_1 n_2 \bar n_2 n_3 \bar n_3} \na^\tbzero{}^{n} \o^\tbzero_i{}^{\bar m_1 m_1} \na^\tbzero{}^{\bar m} \o^\tbzero_j{}^{\bar n_1 n_1} R^\tbzero{}^{\bar m_2 m_2 \bar n_2 n_2} R^\tbzero{}^{\bar m_3 m_3 \bar n_3 n_3}\ .
\ea
The observation that both equations \eqref{first_order_warp} and \eqref{second_order_warp} can be represented as total derivatives in the internal space reflects the topological nature of the terms appearing in \eqref{warpfactoreq}.

It turns out that the tensor $Z_{m \bar m n \bar n}$ given in \eqref{def-Z} plays 
a central role in the following and is related 
to the key topological quantities on $Y_4$. It satisfies the identities
\ba
Z_{m \bar m n \bar n} &= Z_{n \bar m m \bar n} = Z_{m \bar n n \bar m}\ , &
\na^\tbzero{}^{m} Z_{m \bar m n \bar n} &= \na^\tbzero{}^{\bar m} Z_{m \bar m n \bar n}  = 0\ .
\ea
It is related to the third Chern-form $c_3^\tbzero{}$ via
\ba \label{def-Z_plain}
Z_{m \bar m} &= i 2 Z_{m \bar m n}{}^n =  \fr12 (*^\tbzero c_3^\tbzero{})_{m \bar m}\, , \nn
\ea
\vspace{-0.5cm}
\ba 
Z &= i 2 Z_{m}{}^m  = *^\tbzero( J^\tbzero{} \we c_3^\tbzero{})\, ,  & 
*^\tbzero (c_3^\tbzero{} \wedge \o^\tbzero_i ) & = - 2 Z_{m \bar n}\o^\tbzero_i{}^{\bar n m}\, , 
 \ea
and yields the fourth Chern-form $c^\tbzero_4$ by contraction with the Riemann tensor as
\ba 
Z_{m \bar m n \bar n} R^\tbzero{}^{\bar m m \bar n n} &= *^\tbzero c^\tbzero_4\ .
\ea
We note that $Y_{i j m \bar n} $ is also related to $Z_{m \bar m n \bar n}$ upon integration as
\ba
\int_{Y_4} Y_{ij}{}_m{}^m *^\tbzero 1= - \fr16 \int_{Y_4} (  i Z_{m \bar n} \o^\tbzero_{i}{}^{\bar r m} \o^\tbzero_{j}{}^{\bar n}{}_{\bar r } + 2  Z_{m \bar n r \bar s} \o_i^\tbzero{}^{\bar n m} \o_j^\tbzero{}^{\bar s r} ) *^\tbzero1 \, , 
\ea
where the right hand side represents the same linear combination that will be relevant in \eqref{NablaGContributions}. 
We will see in the next section that the three-dimensional effective action contains 
the various contractions of $Z_{m \bar m n \bar n} $. Interestingly, the analog quantity on 
Calabi-Yau threefolds has played a key role in the analysis of \cite{Katmadas:2013mma}.

\section{The three-dimensional effective action} \label{3Daction}

In this section we derive the three-dimensional effective action for the scalar and vector fields 
introduced in section \ref{Perturbations_back}. The kinetic terms for the K\"ahler structure deformations 
and vector fields will be discussed. In a flux background also 
Chern-Simons terms are induced and will be included in our analysis.\footnote{Note that these 
terms are topological in nature and key in the study of chiral F-theory spectra and anomalies \cite{Grimm:2011fx,Cvetic:2012xn}.} 
We also study a non-trivial field-dependent scaling symmetry of the kinetic terms, which 
involves a rescaling of the warp-factor. 
Some of the technical details of the performed reduction are supplemented in appendix \ref{ReductionResults}.

Having identified the background of eleven-dimensional action in section \ref{action+solution} and 
a set of perturbations in section \ref{Perturbations_back}  we are now in a position to derive the three-dimensional 
effective action using a dimensional reduction. To systematically approach this task
we will consider an expansion up to second order in the scalar fluctuations $\d v^i$ and vectors $A^i$.
Furthermore, we will restrict our analysis to terms with only   
two external space derivatives and only retain terms up to order $\eppr^2$.

For the convenience of the reader we begin by summarising the full ansatz that we will use in the reduction. The 
perturbed eleven-dimensional metric takes the form  
\ba \label{ReductionAnsatz1}
 d \wh s^2 =& e^{- \fr{512}{3} \eppr^2 (Z | + \pa_i Z | \d v^i + \fr12 \pa_i \pa_j Z | \d v^i \d v^j )  } \Big[ e^{-2 \eppr^2 (W^\tbtwo | + \pa_i W^\tbtwo | \d v^i + \fr12 \pa_i \pa_j W^\tbtwo | \d v^i \d v^j ) }   g_{\m\n} dx^\m dx^\n  \nn \\
 & +  2 e^{  \eppr^2 (W^\tbtwo | + \pa_i W^\tbtwo | \d v^i + \fr12 \pa_i \pa_j W^\tbtwo | \d v^i \d v^j )}  \big( g^\tbzero_{m \bar n} + \o_i^\tbzero{}_{m \bar n} d v^i  \nn \\
 & + \eppr^2 \pa_{m} \pa_{\bar n}  ( \til F | + \r_i^\tb{s} \d v^i + \pa_i \til F | \d v^i + \fr12 \pa_i \pa_j \til F | \d v^i \d v^j )   \big) dy^m dy^\bar n \Big]\ + \cO(\eppr^3) + \cO(\d v^i{}^3)\ , 
\ea
while the perturbed M-theory four-form field strength is given by 
\ba \label{ReductionAnsatz2}
 \wh G  =& \eppr G^\tbone + F^i \we \o_i^\tbzero +  \eppr^2 F^i \we \pa \bar \pa \r_i^\tb{v}  \nn  \\
     &+  *_3 1 \we d e^{ - 3 \eppr^2 (W^\tbtwo | + \pa_i W^\tbtwo | \d v^i + \fr12 \pa_i \pa_j W^\tbtwo | \d v^i \d v^j ) } + \cO(\eppr^3) + \cO(\d v^i{}^3) \, . 
\ea
The rather involved form of this ansatz reflects the fact that the quantities present are expanded in both $\eppr$ and $\d v^i$.
Recall that the symbol $ |$ means evaluation at $\d v^i = 0$, $\partial_i$ are derivatives 
with respect to $v^i$, and $\pa_{m}$, $\pa_{\bar n}$ are space-time derivatives in the lowest-order complex structure of the 
internal manifold. 

The quantities $Z |,\ \pa_i Z |,\ \pa_i \pa_j Z |$ are directly evaluated by using the definition of 
$Z$ given in \eqref{def-Z_plain}. Similarly one proceeds with the derivatives of $\til F = * ( J \we J \we F) $ given in \eqref{g2Expression}.
In contrast, since the warp-factor $W^\tbtwo$ is only known as a solution to the warp-factor equation \eqref{warpfactoreq}
one would have to apply \eqref{first_order_warp} and \eqref{second_order_warp} to 
determine $\pa_i W^\tbtwo |$ and $\pa_i \pa_j W^\tbtwo |$. It turns out to be sufficient, however, to keep 
$\pa_i W^\tbtwo |$ and $\pa_i \pa_j W^\tbtwo |$ throughout the analysis. Remarkably, we will find that 
all contributions involving $\pa_i \pa_j W^\tbtwo |$ precisely cancel, while the first derivatives $\pa_i W^\tbtwo |$
appear in the correct way to ensure the presence of a $v^i$-dependent scaling symmetry involving the warp-factor.
Before turning to the derivation, let us also note that one may include compensators in the effective action along the lines of the discussion presented in \cite{Gray:2003vw,Giddings:2005ff,Douglas:2008jx}. However these do not change the effective action at the studied order. 

In this subsection we only discuss the kinetic terms that are present in the reduction. The reduction process is quite lengthy and makes use of the intermediate results listed in appendix \ref{ReductionResults}. One inserts the ansatz \eqref{ReductionAnsatz1}, \eqref{ReductionAnsatz2} into the eleven-dimensional action \eqref{complete_11Daction}. The dimensional reduction requires numerous partial integrations and uses multiple Schouten and Bianchi identities, which was only possible by using a computer algorithm. 
Our goal was to represent all three-dimensional terms using the higher-curvature tensor $Z_{m\bar m n\bar n}$ introduced in 
\eqref{def-Z}.
Combining all terms of the computation we find the action 
\ba \label{def-Skin}
S_\text{kin} = S_\text{kin}^\tbzero + \eppr\, S_\text{CS}^\tbone+ \eppr^2\, S_\text{kin}^\tbtwo\ , 
\ea
where at zeroth order one has
\ba \label{zeroth_order_result}
S^\tbzero_\text{kin} =  \fr{1}{2 \k_{11}} \int_{\cM_3} \bls &\Omega^\tbzero R *1  + d \d v^i \we * d \d v^j   \int_{Y_4}  \Big( \fr12  \o^\tbzero_{i m \bar n}  \o^\tbzero_{j }{}^{\bar n m}  - \o^\tbzero_{i m}{}^m  \o^\tbzero_{j n}{}^n  \Big) *^\tbzero 1 \nn \\[.1cm]
  &  +  \fr12  F^i \we * F^j \int_{Y_4} \o^\tbzero_{i m \bar n}\o^\tbzero_{j}{}^{\bar n m}  * ^\tbzero1 \brs \ , 
\ea
while at first order one finds the Chern-Simons terms 
\ba
  S^\tbone_\text{CS} = \fr{1}{2 \k_{11}} \int_{\cM_3}  \Theta_{ij}  A^j \we F^i \ ,\qquad \quad
     \Theta_{ij} = \frac{1}{2} \alpha \int_{Y_4} \o^\tbzero_i \wedge \o^\tbzero_j \wedge G^\tbone \, , 
\ea
and at second order 
\ba \label{second_order_result}
S^\tbtwo_\text{kin} =&   \fr{1}{2 \k_{11}} \int_{\cM_3}  \bls \Omega^\tbtwo R *1 +  d \d v^i \we * d \d v^j   \int_{Y_4}   \Big(  3 i \pa_i W^\tbtwo| \o^\tbzero_{j m}{}^m  + 3W^\tbtwo \big( \fr12 \o^\tbzero_{i m \bar n}  \o^\tbzero_{j }{}^{\bar n m}  - \o^\tbzero_{i m}{}^m  \o^\tbzero_{j n}{}^n \big)   & \nn\\
  & \qquad \qquad \qquad \qquad  -768 Z  \o^\tbzero_{i m}{}^m\o^\tbzero_{j n}{}^n +3072 i Z_{m \bar n}  \o^\tbzero_{i}{}^{\bar n m} \o^\tbzero_{j s}{}^s  +3072  Z_{m \bar n r \bar s} \o^\tbzero_i{}^{\bar n m} \o^\tbzero_j{}^{\bar s r } \Big)*^\tbzero 1\nn \\
  &\qquad \qquad \ \ +    F^i \we * F^j  \int_{Y_4}  \Big( (\fr32 W^\tbtwo  + 256  Z)  \o^\tbzero_{i m \bar n} \o^\tbzero_{j}{}^{ \bar n m } + 192 (-7 + a_1)i Z_{m \bar n} \o^\tbzero_{i}{}^{\bar r m} \o^\tbzero_{j}{}^{\bar n}{}_{\bar r } \nn \\
 &\qquad \qquad  \qquad \qquad \qquad \quad \ \ + 384 (1 + a_1)  Z_{m \bar n r \bar s} \o^\tbzero_i{}^{\bar n m} \o^\tbzero_j{}^{\bar s r} \Big) *^\tbzero1\brs \ .
\ea
Here we have abbreviated 
\ba
  \Omega^\tbzero =& \int_{Y_4} \bls 1  + i   \d v^i  \, \o^\tbzero_{i m}{}^{m}  +  \fr12  \d v^i \d v^j  ( \o^\tbzero_{i m \bar n}  \o^\tbzero_{j }{}^{\bar n m} -  \o^\tbzero_{i m}{}^m  \o^\tbzero_{j n}{}^n ) \brs *^\tbzero 1\ , \nn \\
  \Omega^\tbtwo =&  \int_{Y_4}  \bls  3 W^\tbtwo   +   3 \d v^i  \big( \pa_i W^\tbtwo| + i \o^\tbzero_{i m}{}^m  W^\tbtwo\big) 
  +   \d v^j \d v^i  \Big(  \fr32 \pa_i\pa_j W^\tbtwo| 
    \nn\\
& \qquad
   + 3 i \o^\tbzero_{i m}{}^m \pa_j W^\tbtwo| + \fr32 W^\tbtwo \big(   \o^\tbzero_{i m \bar n}  \o^\tbzero_{j }{}^{\bar n m} - \o^\tbzero_{i m}{}^m  \o^\tbzero_{j n}{}^n\big) \Big) \brs  *^\tbzero 1\, . 
\ea
A few comments are in order. Firstly, we show in appendix \ref{ReductionResults} that 
among all the terms in \eqref{s18term_exp} only $\cA$, $\cZ_1$ an $\cZ_2$ contribute, while $\cZ_3$ to $\cZ_6$ vanish 
identically. This implies that the result should depend on two unknown parameters $a_1,a_2$ that 
appear in \eqref{s18term_exp}. It turns out that for the choice $a_1=a_2$ the result simplifies significantly 
and only depends on $Z_{m\bar m n\bar n}$ as is equally true for the reduction of all 
other term in the eleven-dimensional action \eqref{complete_11Daction}. We therefore have chosen 
$a_1=a_2$ in \eqref{second_order_result}. Secondly, we note that, as already mentioned before, 
the scalar functions $\til F$, $\r^\tb{s}_i$ and $\r^\tb{v}_i$ have totally dropped out of this expression. 
This justifies the use of $\text{dim}(H^{1,1}(Y_4))$ deformations $\d v^i$ and vectors $A^i$.

The action \eqref{second_order_result} still depends on $\pa_i \pa_j W^\tbtwo$, however, only through the coefficient 
of the three-dimensional Einstein-Hilbert term. We now wish to Weyl rescale this action to bring it to the Einstein frame
and show that this dependence actually drops. From \eqref{3dweyl} one finds that one needs to redefine the external metric by $g_{\mu \nu} \rightarrow g'_{\mu\nu} = \Omega^{-2} g_{\mu\nu}$ for
\ba
\Omega =  \Omega^\tbzero + \alpha^2 \, \Omega^\tbtwo\ .
\ea
Performing the Weyl rescaling we find that the kinetic terms displayed in \eqref{zeroth_order_result} and \eqref{second_order_result} become
\ba
S^\tbzero_\text{kin}  =   \fr{1}{2 \k_{11}} \int_{\cM_3} \bls & R *1 +    d \d v^i \we * d \d v^j  \frac{1}{\cV_0}  \int_{Y_4} ( \fr12\o^\tbzero_{i m \bar n} \o^\tbzero_{j}{}^{ \bar n m } +  \o^\tbzero_{i m}{}^m \o^\tbzero_{j n}{}^n ) *^\tbzero1 &\nn\\
&+  F^i \we * F^j  \frac{\cV_0}{2} \int_{Y_4} \o^\tbzero_{i m \bar n} \o^\tbzero_{j}{}^{ \bar n m } *^\tbzero1 \brs\ , &
\ea
and 
\ba
S^\tbtwo_\text{kin} = &  \fr{1}{2 \k_{11}} \int_{\cM_3}  \bls d \d v^i \we * d \d v^j   \bigg(  \frac{1}{\cV_0} \int_{Y_4} \Big(- 9 i \pa_i W^\tbtwo | \o^\tbzero_{j m}{}^m  + \fr32 W^\tbtwo |   \o^\tbzero_{i m \bar n}  \o^\tbzero_{j }{}^{\bar n m}  
& \nn\\
  & -   768 Z  \o^\tbzero_{i m}{}^m\o^\tbzero_{j n}{}^n  
  + 3072 i Z_{m \bar n}  \o^\tbzero_{i}{}^{\bar n m} \o^\tbzero_{i s}{}^s 
   +3072  Z_{m \bar n r \bar s} \o_i^{\tbzero \bar n m} \o_j^{\tbzero  \bar s r }  \Big)*^\tbzero 1 & \nn \\
 & -   \frac{1}{\cV_0^2}  \int_{Y_4} \fr32 W^\tbtwo |*^\tbzero1 \int_{Y_4}  \o^\tbzero_{i m \bar n}  \o^\tbzero_{j }{}^{\bar n m} *^\tbzero1 \bigg) &\nn\\
&  +  F^i \we * F^j  \bigg( \cV_0 \int_{Y_4}  \Big( (\fr32 W^\tbtwo | + 256  Z)  \o^\tbzero_{i m \bar n} \o^\tbzero_{j}{}^{ \bar n m } + 192 (-7 + a_1)i Z_{m \bar n} \o^\tbzero_{i}{}^{\bar r m} \o^\tbzero_{j}{}^{\bar n}{}_{\bar r }&\nn\\ 
&  + 384 (1 + a_1)  Z_{m \bar n r \bar s} \o_i^{\tbzero \bar n m} \o_j^{\tbzero \bar s r} \Big) *^\tbzero 1 +  \int_{Y_4}  \fr32 W^\tbtwo | *^\tbzero 1 \int_{Y_4} \o^\tbzero_{i m \bar n} \o^\tbzero_{j}{}^{ \bar n m } *^\tbzero 1 \bigg) \brs\, ,  &
\ea
where here we have introduced the zeroth-order volume 
\ba
\cV_0 = \int_{Y_4} *^\tbzero 1   \ .
\ea

The warp-factor dependence can be nicely captured by introducing the warped volume and warped metric
\ba
\cV_W =& \int_{Y_4} e^{3 \alpha^2 W^\tbtwo} *^\tbzero 1      \ , &
G^W_{ij} =& \frac{1}{2 \cV_W} \int_{Y_4} e^{3 \alpha^2 W^\tbtwo} \o^\tbzero_{i} \wedge *^\tbzero \o^\tbzero_{j} \, , 
\ea
which at zeroth order in $\alpha$ reduce to $\cV_0$ and 
$G_{ij} = \frac{1}{2 \cV_0} \int_{Y_4} \o^\tbzero_{i} \wedge *^\tbzero \o^\tbzero_{j} $. 
We also introduce 
\ba
  K_i^W =& i \cV_W  \, \o^\tbzero_{i m}{}^m + \frac{9}{2} \alpha^2 \int_{Y_4} \pa_i W^\tbtwo| *^\tbzero 1\ ,
\ea
which at lowest order simply reduces to $ K_i = i \cV_0  \, \o^\tbzero_{i m}{}^m= \tfrac{1}{3!} \int_{Y_4}  \o^\tbzero_{i} \wedge  J^\tbzero \wedge J^\tbzero  \wedge J^\tbzero $.
With these definitions one rewrites the action \eqref{def-Skin} for all kinetic terms into the form 
\ba
S_\text{kin}  =&   \fr{1}{2 \k_{11}} \int_{\cM_3} \bls  R *1 -   (G^W_{ij} +\cV_W^{-2} K_i^W  K_j^W )d v^i \we * d v^j  
    -  \cV^{2}_W G_{ij}^W F^i \we * F^j + \Theta_{ij} A^i \wedge F^i  & \nn \\
     & - dv^i \we * d v^j    \frac{\alpha^2}{\cV_0} \int_{Y_4} \Big(768 Z  \o^\tbzero_{i m}{}^m\o^\tbzero_{j n}{}^n   - 3072 i Z_{m \bar n}  \o^\tbzero_{i}{}^{\bar n m} \o^\tbzero_{j s}{}^s  -3072  Z_{m \bar n r \bar s} \o_i^{\tbzero \bar n m} \o_j^{\tbzero  \bar s r }  \Big)*^\tbzero 1  &\nn\\
&  +  F^i \we * F^j   \alpha^2 \cV_0 \int_{Y_4} \Big( 256  Z \o^\tbzero_{i m \bar n} \o^\tbzero_{j}{}^{ \bar n m } + 192 (-7 + a_1)i Z_{m \bar n} \o^\tbzero_{i}{}^{\bar r m} \o^\tbzero_{j}{}^{\bar n}{}_{\bar r }  &\nn\\ 
&  \qquad \qquad \quad + 384 (1 + a_1)  Z_{m \bar n r \bar s} \o_i^{\tbzero \bar n m} \o_j^{\tbzero \bar s r} \Big) *^\tbzero1  \brs\ , &  
\ea
where we have replaced $d \d v^i$ directly with $d v^i$.
Expanding to order $\alpha^2$ one indeed recovers the above result. 

It is interesting to observe that the three-dimensional effective action permits a scaling symmetry 
involving the rescaling of the warp-factor. 
 We begin by noting that the eleven-dimensional background ansatz given in subsection \ref{11Dsolution}
 has a symmetry under which
\ba
W^\tbtwo &\ra W^\tbtwo + \Lambda^\tbtwo  \ ,&
g_{m \bar n} &\ra e^{- \eppr^2   \Lambda^\tbtwo}  g_{m \bar n}\ , &
g_{\m \n} &\ra e^{2 \eppr^2 \Lambda^\tbtwo}  g_{\m \n} \ ,
\ea
for $\Lambda^\tbtwo = \Lambda^\tbtwo(x^\m) $. 
This can be extended to a symmetry of the perturbed background \eqref{ReductionAnsatz1} and \eqref{ReductionAnsatz2} by requiring that 
\beq
   v^i \ra e^{ - \eppr^2 \Lambda^\tbtwo} v^i\ .
\eeq
This then implies that 
\ba
d v^i \ra e^{ - \eppr^2\Lambda^\tbtwo} d v^i -  \eppr^2  v^i  \, \pa_j \Lambda^\tbtwo d v^j \ ,
\ea
if we further restrict $\Lambda^\tbtwo = \Lambda^\tbtwo(v^i) $.
When the reduction is performed this the becomes a symmetry of the effective action before the Weyl rescaling to move to the Einstein frame is performed. When the rescaling is performed the value of $\Omega$ in $g_{\mu \nu} \rightarrow g'_{\mu\nu} = \Omega^{-2} g_{\mu\nu}$ transforms as $\Omega \ra e^{- \eppr^2 W^\tbtwo} \O$ so that the rescaled metric does not transform. The final form of the effective action coming from the dimensional reduction is then invariant under the symmetry
\ba 
\label{symmetry}
&W^\tbtwo \ra W^\tbtwo + \Lambda^\tbtwo  \ , & 
&   v^i \ra e^{ - \eppr^2 \Lambda^\tbtwo} v^i \ .
\ea
We note that the $\pa_i W^\tbtwo$ terms in the $\d v^i$ kinetic terms are key to ensuring the symmetry of the action for $\Lambda^\tbtwo$ as a function of $v^i$, as they covariantize the derivatives which appear in the reduction. 
Indeed, this symmetry can be made manifest by introducing a covariant derivative for $v^i$. Furthermore we note that if we make the choice $a_1 = 7$ then using the definitions,
\ba
G^T_{i j} & = G^W_{i j} + 256 \fr{1}{\cV_0^2} \int_{Y_4} Z *^\tbzero 1  \int_{Y_4} \o_i^\tbzero{}_{m \bar n } \o_j^\tbzero{}^{\bar n m} *^\tbzero 1 \nn \hspace{-20cm} & \\& 
-  256 \fr{1}{\cV_0} \int_{Y_4} \bls Z \o_i^\tbzero{}_{m \bar n } \o_j^\tbzero{}^{\bar n m}  + 12 Z_{m \bar n r \bar s} \o_j^\tbzero{}^{\bar n m}   \o_i^\tbzero{}^{\bar s r} \brs *^\tbzero 1 \nn \,  , \hspace{-20cm} &\\
K_i^T & = K_i + \eppr^2 \int_{Y_4}  \bls \fr{1}{3!}  ( 3  W^\tbtwo  - 128 Z )   J^\tbzero \we J^\tbzero \we J^\tbzero  \we \o^\tbzero_i - 1536  Z_{m \bar n} \o_i^\tbzero{}^{\bar n m}  *^\tbzero 1 \brs \, ,  \nn \hspace{-20cm}& \\
\cV_T & =  \cV_W + \eppr^2 256 \int_{Y_4} Z *^\tbzero 1 \, , & 
D v^i & = d v^i + \eppr^2 v^i \pa_j W^\tbtwo d v^j \, , 
\ea
the action takes the simple form\footnote{Note that in making this match we have used that 
\ba
\int_{\cM_3} d v^i \we * dv^j \fr{1}{\cV_0} \int_{Y_4} Z \o^\tbzero_i \we *^\tbzero \o^\tbzero_j &=\int_{\cM_3} d v^i \we * dv^j \fr{1}{\cV_0^2} \int_{Y_4} Z ^\tbzero*1 \int_{Y_4} \o^\tbzero_i \we *^\tbzero \o^\tbzero_j \nn \\
\int_{\cM_3} d v^i \we * dv^j \fr{1}{\cV_0}\int_{Y_4} W \o_i^\tbzero \we *^\tbzero \o_j^\tbzero &= \int_{\cM_3} d v^i \we * dv^j \fr{1}{\cV_0^2} \int_{Y_4} W *^\tbzero1 \int_{Y_4} \o^\tbzero_i \we *^\tbzero \o^\tbzero_j
\ea
which can be demonstrated by taking using integration by parts in the external space. } 
\ba
\label{FinalResult}
 S_\text{kin}  =&  \fr{1}{2 \k_{11}} \int_{\cM_3} \bls  R *1 -   (\cG^T_{ij} +\cV_T^{-2} \cK_i^T  \cK_j^T ) D v^i \we * D v^j  
    -  \cV^{2}_T \cG_{ij}^T F^i \we * F^j + \Theta_{ij} A^i \wedge F^i \brs\, .
\ea
Where now it is clear that under  \eqref{symmetry}
\ba
\cG^T_{ij}  & \rightarrow e^{2 \alpha^2 \Lambda^\tbtwo} \cG^T_{ij} \, , &  
\cV_T & \rightarrow e^{- \alpha^2 \Lambda^\tbtwo} \cV_T\, , & 
\cK_i^T & \ra e^{2 \alpha^2 \Lambda^\tbtwo} K^T_i\, , &  
D v^i & \ra e^{- \alpha^2 \Lambda^\tbtwo} D v^i \,, & 
\ea
 so that the action \eqref{FinalResult} is invariant.  

Finally, let us briefly discuss the potential of the three-dimensional effective theory. 
It is well-known that it contains a flux-dependent part given by \cite{Haack:1999zv,Haack:2001jz} 
\ba \label{Spot}
S_\text{pot}  = \fr{\alpha^2}{4 \k_{11}} \int_{\cM_3} *1 \int_{Y_4} (G^\tbone \we * G^\tbone - G^\tbone\we G^\tbone) \, , 
\ea
in which the internal Hodge star is evaluated in the perturbed zeroth-order metric \eqref{flucg} which sees the full $v^i$. 
This term is responsible for imposing self-duality of $G^\tbone$ in the vacuum. 
With our restriction to K\"ahler deformations this implies that $G^\tbone$ 
remains primitive with respect to the perturbed metric \eqref{flucg} for massless 
fluctuations, i.e.~the ones satisfying \eqref{flucprimitive}.
We can easily see that warping or higher-curvature
corrections which multiply this result will yield corrections that are higher than order $\eppr^2$ and 
thus cannot be reliably analyzed using our ansatz. 
Furthermore, we propose that at order $\alpha^2$ there are no terms added to \eqref{Spot} 
that are only dependent on the warping and internal space higher-curvature terms, 
as the background we analyse is invariant under the perturbations we consider 
as long as $G^\tbone$ remains self-dual. This will be demonstrated in \cite{PaperPartII}.

Let us close by noting that in a next step one has to bring the action into standard $\cN=2$ form
and determine a kinetic potential and the correct $\cN=2$ coordinates. 
This will be done in the second part of this paper \cite{PaperPartII}.


\begin{appendix}
\vspace{2cm} 
\noindent {\bf \LARGE Appendix}

\section{Conventions, definitions, and identities} \label{Conventions}

In this work we denote the eleven-dimensional space indices by capital Latin letters $M,N,R = 0,\ldots,10$,
the external  ones by  $\mu,\nu = 0,1,2$, and the internal complex ones by $m,n,p=1,...,4$ and $\bar m, \bar n,\bar p =1, \ldots,4$.
Eleven-dimensional quantities for which the indices are raised and lower with the total space metric carry a hat, for example the 
M-theory three-form is denoted by $\wh G$. 
Furthermore, the convention for the totally 
anti-symmetric tensor in Lorentzian space in an orthonormal frame is $\epsilon_{012...10} = \epsilon_{012}=+1$. 
The epsilon tensor in $d$ dimensions then satisfies
\ba
\epsilon^{R_1\cdots R_p N_{1 }\ldots N_{d-p}}\epsilon_{R_1 \ldots R_p M_{1} \ldots M_{d-p}} &= (-1)^s (d-p)! p! 
\delta^{N_{1}}{}_{[M_{1}} \ldots \delta^{N_{d-p}}{}_{M_{d-p}]} \,, 
\ea
where  $s=0$ if the metric has Riemannian signature and $s=1$ for a Lorentzian metric.

We adopt the following conventions for the Christoffel symbols and Riemann tensor 
\ba
\G^R{}_{M N} & = \fr12 g^{RS} ( \pa_{M} g_{N S} + \pa_N g_{M S} - \pa_S g_{M N}  ) \, , &
R_{M N} & = R^R{}_{M R N} \, , \nn \\
R^{M}{}_{N R S} &= \pa_R \G^M{}_{S N}  - \pa_{S} \G^M{}_{R N} + \G^M{}_{R  T} \G^T{}_{S N} - \G^M{}_{ST} \G^T{}_{R N} \,, &
R & = R_{M N} g^{M N} \, , 
\ea
with equivalent definitions on the internal and external spaces. 

The terms $\wh t_8 \wh t_8 \wh R^4$ and $\wh t_8 \wh t_8 \wh G^2 \wh R^3$ in \eqref{M-TheoryAction_1} and \eqref{M-TheoryAction_2} require the definition
\ba \label{def-t8}
\hat t_8^{N_1\dots N_8}   &= \fr{1}{16} \big( -  2 \left(   \wh g^{ N_1 N_3  }\wh g^{  N_2  N_4  }\wh g^{ N_5   N_7  }\wh g^{ N_6 N_8  } 
 + \wh g^{ N_1 N_5  }\wh g^{ N_2 N_6  }\wh g^{ N_3   N_7  }\wh g^{  N_4   N_8   }
 +  \wh g^{ N_1 N_7  }\wh g^{ N_2 N_8  }\wh g^{ N_3   N_5  }\wh g^{  N_4 N_6   }  \right) \nn \\
 & \quad +
 8 \left(  \wh g^{  N_2     N_3   }\wh g^{ N_4    N_5  }\wh g^{ N_6    N_7  }\wh g^{ N_8   N_1   } 
  +\wh g^{  N_2     N_5   }\wh g^{ N_6    N_3  }\wh g^{ N_4    N_7  }\wh g^{ N_8   N_1   } 
  +   \wh g^{  N_2     N_5   }\wh g^{ N_6    N_7  }\wh g^{ N_8    N_3  }\wh g^{ N_4  N_1   } 
\right) \nn \\
& \quad - (N_1 \lra N_2) -( N_3 \lra N_4) - (N_5 \lra N_6) - (N_7 \lra N_8) \big) \,. 
\ea
In order to discuss the term $\wh s_{18}$ appearing in \eqref{M-TheoryAction_3} and \eqref{s18term_exp}
we introduce the basis 
\vspace*{-1cm}
\begin{center}
\resizebox{1\textwidth}{!}{
\hspace*{-.1cm}\begin{minipage}[c]{\textwidth}
\ba
B_{1}  &=  \wh R_\sm{N_{1}N_{2} N_{3}N_{4}} \wh R_\sm{N_{5}N_{6}N_{7}N_{8}}             \wh \na^\sm{N_{5}}\wh G^\sm{N_{1}N_{7}N_{8}}{}_\sm{N_{9}}    \wh \na^\sm{N_{3}}\wh G^\sm{N_{2}N_{4}N_{6}N_{9}} \,,  &
B_{13} &= \wh R_\sm{N_{1}N_{2}N_{3}N_{4}} \wh R_\sm{N_{5}}{}^\sm{N_{1}}{}_\sm{N_{6}}{}^\sm{N_{3}} \wh \na_\sm{N_{9}}\wh G^\sm{N_{2}N_{6}}{}_\sm{N_{7}N_{8}} \wh \na^\sm{N_{9}}\wh G^\sm{N_{4}N_{5}N_{7}N_{8}} \,, \nn \\
B_{2}  &=  \wh R_\sm{N_{1}N_{2}N_{3}N_{4}} \wh R_\sm{N_{5}N_{6}N_{7}N_{8}}             \wh \na^\sm{N_{5}}\wh G^\sm{N_{1}N_{3}N_{7}}{}_\sm{N_{9}}    \wh \na^\sm{N_{8}}\wh G^\sm{N_{2}N_{4}N_{6}N_{9}} \,, &
B_{14} &= \wh R_\sm{N_{1}N_{2}N_{3}N_{4}} \wh R_\sm{N_{5}}{}^\sm{N_{1}}{}_\sm{N_{6}}{}^\sm{N_{3}} \wh \na_\sm{N_{9}}\wh G^\sm{N_{2}N_{4}}{}_\sm{N_{7}N_{8}} \wh \na^\sm{N_{9}}\wh G^\sm{N_{5}N_{6}N_{7}N_{8}} \,, \nn \\
B_{3}  &=  \wh R_\sm{N_{1}N_{2}N_{3}N_{4}} \wh R_\sm{N_{5}N_{6}N_{7}N_{8}}             \wh \na^\sm{N_{5}}\wh G^\sm{N_{1}N_{3}N_{7}}{}_\sm{N_{9}}    \wh \na^\sm{N_{6}}\wh G^\sm{N_{2}N_{4}N_{8}N_{9}} \,, &
B_{15} &= \wh R_\sm{N_{1}N_{2}N_{3}N_{4}} \wh R_\sm{N_{5}}{}^\sm{N_{1}}{}_\sm{N_{6}}{}^\sm{N_{3}} \wh \na^\sm{N_{2}}\wh G^\sm{N_{6}}{}_\sm{N_{7}N_{8}N_{9}} \wh \na^\sm{N_{5}}\wh G^\sm{N_{4}N_{7}N_{8}N_{9}} \,, \nn \\
B_{4}  &=  \wh R_\sm{N_{1}N_{2}N_{3}N_{4}} \wh R_\sm{N_{5}N_{6}N_{7}N_{8}}             \wh \na_\sm{N_{9}}\wh G^\sm{N_{3}N_{4}N_{7}N_{8}}         \wh \na^\sm{N_{6}}\wh G^\sm{N_{9}N_{1}N_{2}N_{5}} \,,&
B_{16} &= \wh R_\sm{N_{1}N_{2}N_{3}N_{4}} \wh R_\sm{N_{5}}{}^\sm{N_{1}}{}_\sm{N_{6}}{}^\sm{N_{3}} \wh \na^\sm{N_{2}}\wh G^\sm{N_{4}}{}_\sm{N_{7}N_{8}N_{9}} \wh \na^\sm{N_{5}}\wh G^\sm{N_{6}N_{7}N_{8}N_{9}} \,,\nn \\
B_{5}  &=  \wh R_\sm{N_{1}N_{2}N_{3}N_{4}} \wh R_\sm{N_{5}N_{6}N_{7}}{}^\sm{N_{4}}        \wh \na^\sm{N_{1}}\wh G^\sm{N_{2}N_{3}}{}_\sm{N_{8}N_{9}}    \wh \na^\sm{N_{5}}\wh G^\sm{N_{6}N_{7}N_{8}N_{9}} \,, &
B_{17} &= \wh R_\sm{N_{1}N_{2}N_{3}N_{4}} \wh R_\sm{N_{5}}{}^\sm{N_{1}}{}_\sm{N_{6}}{}^\sm{N_{3}} \wh \na^\sm{N_{2}}\wh G^\sm{N_{5}}{}_\sm{N_{7}N_{8}N_{9}} \wh \na^\sm{N_{4}}\wh G^\sm{N_{6}N_{7}N_{8}N_{9}} \,,\nn \\
B_{6}  &=  \wh R_\sm{N_{1}N_{2}N_{3}N_{4}} \wh R_\sm{N_{5}N_{6}N_{7}}{}^\sm{N_{4}}        \wh \na^\sm{N_{1}}\wh G^\sm{N_{2}N_{5}}{}_\sm{N_{8}N_{9}}    \wh \na^\sm{N_{3}}\wh G^\sm{N_{6}N_{7}N_{8}N_{9}} \,, &
B_{18} &= \wh R_\sm{N_{1}N_{2}N_{3}N_{4}} \wh R_\sm{N_{5}}{}^\sm{N_{1}}{}_\sm{N_{6}}{}^\sm{N_{3}} \wh \na_\sm{N_{9}}\wh G^\sm{N_{5}N_{6}}{}_\sm{N_{7}N_{8}} \wh \na^\sm{N_{4}}\wh G^\sm{N_{2}N_{7}N_{8}N_{9}} \,,\nn  \\
B_{7}  &=  \wh R_\sm{N_{1}N_{2}N_{3}N_{4}} \wh R_\sm{N_{5}N_{6}N_{7}}{}^\sm{N_{4}}        \wh \na^\sm{N_{1}}\wh G^\sm{N_{2}N_{5}}{}_\sm{N_{8}N_{9}}    \wh \na^\sm{N_{7}}\wh G^\sm{N_{3}N_{6}N_{8}N_{9}} \,,&
B_{19} &= \wh R_\sm{N_{1}N_{2}N_{3}N_{4}} \wh R_\sm{N_{5}N_{6}}{}^\sm{N_{3}N_{4}}        \wh \na_\sm{N_{9}}\wh G^\sm{N_{1}N_{5}}{}_\sm{N_{7}N_{8}}    \wh \na^\sm{N_{9}}\wh G^\sm{N_{2}N_{6}N_{7}N_{8}} \,, \nn \\
B_{8}  &=  \wh R_\sm{N_{1}N_{2}N_{3}N_{4}} \wh R_\sm{N_{5}N_{6}N_{7}}{}^\sm{N_{4}}        \wh \na^\sm{N_{1}}\wh G^\sm{N_{3}N_{5}}{}_\sm{N_{8}N_{9}}    \wh \na^\sm{N_{2}}\wh G^\sm{N_{6}N_{7}N_{8}N_{9}} \,, &
B_{20} &= \wh R_\sm{N_{1}N_{2}N_{3}N_{4}} \wh R_\sm{N_{5}N_{6}}{}^\sm{N_{3}N_{4}}        \wh \na^\sm{N_{1}}\wh G^\sm{N_{5}}{}_\sm{N_{7}N_{8}N_{9}}    \wh \na^\sm{N_{2}}\wh G^\sm{N_{6}N_{7}N_{8}N_{9}} \,, \nn \\
B_{9}  &=  \wh R_\sm{N_{1}N_{2}N_{3}N_{4}} \wh R_\sm{N_{5}N_{6}N_{7}}{}^\sm{N_{4}}        \wh \na^\sm{N_{1}}\wh G^\sm{N_{3}N_{5}}{}_\sm{N_{8}N_{9}}    \wh \na^\sm{N_{6}}\wh G^\sm{N_{2}N_{7}N_{8}N_{9}} \,, &
B_{21} &= \wh R_\sm{N_{1}N_{2}N_{3}N_{4}} \wh R_\sm{N_{5}N_{6}}{}^\sm{N_{3}N_{4}}        \wh \na^\sm{N_{1}}\wh G^\sm{N_{5}}{}_\sm{N_{7}N_{8}N_{9}}    \wh \na^\sm{N_{6}}\wh G^\sm{N_{2}N_{7}N_{8}N_{9}} \,, \nn \\
B_{10} &= \wh R_\sm{N_{1}N_{2}N_{3}N_{4}} \wh R_\sm{N_{5}N_{6}N_{7}}{}^\sm{N_{4}}        \wh \na_\sm{N_{9}}\wh G^\sm{N_{3}N_{5}N_{7}N_{8}}         \wh \na^\sm{N_{9}}\wh G^\sm{N_{1}N_{2}N_{6}N_{8}} \,, &
B_{22} &= \wh R_\sm{N_{1}N_{2}N_{3}N_{4}} \wh R_\sm{N_{5}}{}^\sm{N_{1}N_{3}N_{4}}        \wh \na^\sm{N_{2}}\wh G_\sm{N_{6}N_{7}N_{8}N_{9}}         \wh \na^\sm{N_{5}}\wh G^\sm{N_{6}N_{7}N_{8}N_{9}} \,, \nn \\
B_{11} &= \wh R_\sm{N_{1}N_{2}N_{3}N_{4}} \wh R_\sm{N_{5}N_{6}N_{7}}{}^\sm{N_{4}}        \wh \na_\sm{N_{8}}\wh G^\sm{N_{1}N_{2}N_{6}}{}_\sm{N_{9}}    \wh \na^\sm{N_{9}}\wh G^\sm{N_{3}N_{5}N_{7}N_{8}}  \,, &
B_{23} &= \wh R_\sm{N_{1}N_{2}N_{3}N_{4}} \wh R_\sm{N_{5}}{}^\sm{N_{1}N_{3}N_{4}}        \wh \na_\sm{N_{9}}\wh G^\sm{N_{2}}{}_\sm{N_{6}N_{7}N_{8}}    \wh \na^\sm{N_{9}}\wh G^\sm{N_{5}N_{6}N_{7}N_{8}} \,, \nn \\
B_{12} &= \wh R_\sm{N_{1}N_{2}N_{3}N_{4}} \wh R_\sm{N_{5}N_{6}N_{7}}{}^\sm{N_{4}}        \wh \na^\sm{N_{3}}\wh G^\sm{N_{5}N_{6}}{}_\sm{N_{8}N_{9}}    \wh \na^\sm{N_{7}}\wh G^\sm{N_{2}N_{1}N_{8}N_{9}} \,, &
B_{24} &= \wh R_\sm{N_{1}N_{2}N_{3}N_{4}} \wh R^\sm{N_{1}N_{2}N_{3}N_{4}}             \wh \na_\sm{N_{5}}\wh G_\sm{N_{6}N_{7}N_{8}N_{9}}         \wh \na^\sm{N_{6}}\wh G^\sm{N_{5}N_{7}N_{8}N_{9}} \,. 
\nn
\ea
\end{minipage}
}
\vspace{-.1cm}
\beq
 \ \label{def-Bi}
\eeq
\end{center}
The contributions to $\wh s_{18} (\wh \na \wh G)^2 \wh R^2 $ are then formed from the linear combinations described in  \eqref{s18term_exp}.

We note that performing a Weyl rescaling of the three dimensional external metric with $g'_{\mu\nu} = \Omega^{-2} g_{\mu\nu}$ one finds that
\ba
\label{3dweyl}
\int_{\cM_3} \Omega R' *'_3 1 = \int_{\cM_3} (R *_3 1 - \frac{2}{\Omega^2} \nabla_\mu \Omega \nabla^\mu \Omega *_3 1 ) \,. 
\ea

Finally we demonstrate that the 3d effective action may be simplified by using the intersection structures
\ba
K_{i j k l}  &= \int_{Y_4} \o^\tbzero_i \we   \o^\tbzero_j \we   \o^\tbzero_k \we   \o^\tbzero_l \, ,& 
K_{i j k}  & = K_{i j k l} v^l \,, & 
K_{i j}  & = \fr12 K_{i j k l} v^k v^l \, , & \nn \\
K_{i }  & = \fr1{3!} K_{i j k l} v^j v^k v^l \, ,& 
\cV_0  & = \fr1{4!} K_{i j k l} v^i v^j v^k v^l \, , & 
\ea 
\section{Results of the dimensional reduction}
\label{ReductionResults}

\subsection{Two derivative terms}

The reduction of the lowest order part of the action \eqref{complete_11Daction} gives the following contribution to the Kinetic terms of the 3d theory
\ba
\label{3dAction1}
S^\tbzero |_\text{kin} =   \fr{1}{2 \k_{11}} \int_{\cM_3}& R *1 \int_{Y_4} \bls  e^{  \eppr^2 (3 W^\tbtwo  - 768 Z)   } \Big( 1  + i \d v^i \o^\tbzero_{ i m }{}^m  + \frac{1}{2} \d v^i \d v^j \left( \o^\tbzero_{i m \bar n}  \o^\tbzero_{j }{}^{\bar n m} - \o^\tbzero_{i m}{}^m \o_{j n}{}^n \right) \Big)  & \nn \\  
 &  + 3 \eppr^2  \d v^i  \pa_iW^\tbtwo |  +    3 i \eppr^2    \d v^i \d v^j \,  \pa_{(i}W^\tbtwo |  \, \o_{j) m}{}^m  +  \fr32 \eppr^2  \d v^i \d v^j   \pa_{i}\pa_{j}W^\tbtwo| 
  +  1536  \eppr^2  \d v^i  Z_{m \bar n} \o^\tbzero_{i}{}^{\bar n m} 
   \nn \\
 &
 + i 768  \eppr^2  Z  \d v^i  \o^\tbzero_{i m }{}^{m} + 384  \eppr^2  Z  \d v^i   \d v^j \o^\tbzero_{i m \bar n } \o^\tbzero_{j}{}^{\bar n m} - 384  \eppr^2  \d v^i  \d v^j Z  \o^\tbzero_{i m }{}^{m} \o^\tbzero_{j n }{}^{n}
 \brs *^\tbzero 1&  \nn \\   
 +  \fr{1}{2 \k_{11}} \int_{\cM_3} & d \d v^i \we * d \d v^j   \int_{Y_4}   \bls     e^{\eppr^2 (3 W^\tbtwo  - 768 Z) } \left( \fr12 \o^\tbzero_{i m \bar n}  \o^\tbzero_{j }{}^{\bar n m} -  \o^\tbzero_{i m}{}^m\o_{j n}{}^n \right)  & \nn \\  
 &  + 3i  \eppr^2 \pa_{(i} W^\tbtwo|  \, \o^\tbzero_{j )m}{}^m 
 +3072 \eppr^2  i Z_{m \bar n}  \o^\tbzero_{i}{}^{\bar n m} \o^\tbzero_{j s}{}^s - 1536  \eppr^2  Z \o^\tbzero_{i  m}{}^{m} \o^\tbzero_{j  n}{}^n  \brs *^\tbzero 1  \ 
 &\nn \\
 + \fr{1}{2 \k_{11}}\fr12 \int_{\cM_3} & F^i \we * F^j \int_{Y_4} e^{  \eppr^2 (  3 W^\tbtwo - 256   Z)} \o^\tbzero_{i m \bar n}\o^\tbzero_{j}{}^{\bar n m} *^\tbzero 1 + \eppr \fr{1}{2 \k_{11}}\int_{\cM_3} F^i \we A^j \int_{Y_4}\fr12  G^\tbone \we  \o^\tbzero_i  \we \o^\tbzero_j \,. &
\ea

It is interesting to note that in these terms the value of $\til F$, $\r_i^\tb{s}$ and  $\r_i^\tb{v}$ drop out of these expressions as they contribute only internal space total derivatives to the 3d effective theory. 

\subsection{Eight derivative terms}

Let us record the reduction of certain higher derivative terms which are used as intermediate results in deriving the effective action \eqref{def-Skin}. These results were computed using the mathematica package xAct and required the use of several internal space total derivative and schouten identities. 
\ba
\int  \wh t_8 \wh t_8 \wh R^4 \wh * 1 |_\text{kin}& =  \fr{1}{2 \k_{11}} \int_{\cM_3}  d \d v^i \we * d \d v^j   \int_{Y_4} 384  \left(   Z   \o^\tbzero_{i m \bar n}{}  \o^\tbzero_{j }{}^{\bar n m} + 4  Z_{m \bar n r \bar s} \o^\tbzero_i{}^{\bar n m} \o^\tbzero_j{}^{\bar s r } \right) *^\tbzero 1\, ,\nn 
\ea
\vspace{-0.5cm}
\ba
 - \fr{1}{24} \int  \wh \e_{11} \wh \e_{11} \wh R^4 \wh * 1 |_\text{kin}& = \fr{1}{2 \k_{11}} \int_{\cM_3} R *1 \int_{Y_4} \Big( 768  Z  - 1536  \d v^i  Z_{m \bar n}\o^\tbzero_{i}{}^{\bar n m} \Big) *^\tbzero 1 \nn  \\ 
& +   \fr{1}{2 \k_{11}} \int_{\cM_3}  d \d v^i \we * d \d v^j   \int_{Y_4} 1536  Z_{m \bar n r \bar s} \o^\tbzero_i{}^{ \bar n m} \o^\tbzero_j{}^{ \bar s r } *^\tbzero 1 \, ,\nn 
\ea
\vspace{-0.5cm}
\ba
\int_{Y_4} 3^2 2^{13}  \wh C \we \wh X_8 |_\text{kin}& = 0 \, . 
\ea

Similarly we note that the reduction of the $\wh G^2 \wh R^3$ terms uses the identities
\ba
- \int  \wh t_8 \wh t_8 \wh G^2 \wh R^3 \wh * 1 |_\text{kin}& =  \fr{1}{2 \k_{11}^2}   
384 \int_{\cM_3} F^i \we * F^j \int_{Y_4} \bls Z \o^\tbzero_{i m \bar n} \o^\tbzero_{j}{}^{ \bar n m } 
\nn \\ & 
- 4i Z_{m \bar n} \o^\tbzero_{i}{}^{\bar r m} \o^\tbzero_{j}{}^{\bar n}{}_{\bar r } - 4 Z_{m \bar n r \bar s} \o^\tbzero{}_i^{ \bar n m} \o^\tbzero{}_j^{ \bar s r } \brs *^\tbzero 1 , \nn 
\ea
\vspace{-0.5cm}
\ba
- \fr{1}{96} \int  \wh \e_{11} \wh \e_{11} \wh G^2 \wh R^3 \wh * 1 |_\text{kin}& =  \fr{1}{2 \k_{11}^2}   
1536 \int_{\cM_3} F^i \we * F^j \int_{Y_4}  Z_{m \bar n r \bar s} \o^\tbzero_i{}^{ \bar n m} \o^\tbzero_j{}^{ \bar s r } *^\tbzero 1 .
\ea

Finally reducing the $( \wh \na \wh G)^2 \wh R^2$ terms in \eqref{complete_11Daction} gives
\ba
\int &\wh s_{18} ( \wh \na \wh G)^2 \wh R^2  \wh * 1 |_\text{kin}
 =  \fr{1}{2 \k_{11}^2}   
\int_{\cM_3} \fr12 F^i \we * F^j \int_{Y_4} \bls 
-96(1 + a_2)\o^\tbzero_i{}^{\bar 
m}{}^{n}\o^\tbzero_j{}^{\bar r}{}^{r}R^\tbzero{}^{ \bar s}{}_{\bar 
r}{}_{n}{}^{s}R^\tbzero{}^{\bar t}{}_{\bar s}{}_{r}{}^{t}R^\tbzero{}_{s}{}_{\bar 
m}{}_{t}{}_{ \bar t} \nn \\ &
- 48(2 + a_1 + 
a_2)\o^\tbzero_i{}^{\bar 
m}{}^{n}\o^\tbzero_j{}^{\bar r}{}^{r}R^\tbzero{}^{ \bar 
s}{}^{s}{}_{t}{}^{u}R^\tbzero{}_{n}{}_{\bar m}{}_{r}{}^{t}R^\tbzero{}_{s}{}_{\bar 
r}{}_{u}{}_{\bar s} 
+ 48(1 + 
a_1)\o^\tbzero_i{}^{\bar 
m}{}^{n}\o^\tbzero_j{}^{\bar r}{}^{r}R^\tbzero{}^{ \bar s}{}_{\bar 
m}{}_{r}{}^{s}R^\tbzero{}^{\bar t}{}_{\bar r}{}_{n}{}^{t}R^\tbzero{}_{s}{}_{\bar 
s}{}_{t}{}_{ \bar t} 
\nn \\ &
+ 48(1 + 
a_2)\o^\tbzero_i{}^{\bar 
m}{}^{n}\o^\tbzero_j{}^{\bar r}{}^{r}R^\tbzero{}^{ \bar s}{}_{\bar 
m}{}_{n}{}^{s}R^\tbzero{}^{\bar t}{}_{\bar r}{}_{r}{}^{t}R^\tbzero{}_{s}{}_{\bar 
s}{}_{t}{}_{ \bar t} 
- 48(2 + a_1 + 
a_2)\o^\tbzero_i{}^{\bar 
m}{}^{n}\o^\tbzero_j{}^{\bar r}{}^{r}R^\tbzero{}^{ \bar s}{}_{\bar 
m}{}_{n}{}_{\bar r}R^\tbzero{}^{\bar t}{}^{s}{}_{r}{}^{t}R^\tbzero{}_{s}{}_{\bar 
s}{}_{t}{}_{ \bar t} 
\nn \\ &
+ 24(1 + 
a_1)\o^\tbzero_i{}^{\bar 
m}{}^{n}\o^\tbzero_j{}^{\bar r}{}^{r}R^\tbzero{}^{ \bar s}{}^{s}{}^{\bar 
t}{}^{t}R^\tbzero{}_{n}{}_{\bar m}{}_{r}{}_{\bar r}R^\tbzero{}_{s}{}_{\bar 
s}{}_{t}{}_{ \bar t} 
+ 48(1 + 
a_2)\o^\tbzero_i{}^{\bar 
m}{}^{n}\o^\tbzero_j{}^{\bar r}{}^{r}R^\tbzero{}^{ \bar 
s}{}^{s}{}_{n}{}^{t}R^\tbzero{}_{r}{}_{\bar s}{}_{s}{}^{u}R^\tbzero{}_{t}{}_{\bar 
m}{}_{u}{}_{\bar r} 
\nn \\ &
+ 48(a_1 - 
a_2)\o^\tbzero_i{}^{\bar 
m}{}^{n}\o^\tbzero_j{}^{\bar r}{}^{r}R^\tbzero{}^{ \bar s}{}_{\bar 
r}{}_{n}{}^{s}R^\tbzero{}_{r}{}^{t}{}_{s}{}^{u}R^\tbzero{}_{t}{}_{\bar m}{}_{u}{}_{\bar 
s} 
- 48(1 + 
a_1)\o^\tbzero_i{}^{\bar 
m}{}^{n}\o^\tbzero_j{}_{n}{}^{r}R^\tbzero{}^{\bar r}{}_{\bar m}{}_{s}{}^{t}R^\tbzero{}^{ \bar 
s}{}^{s}{}_{r}{}^{u}R^\tbzero{}_{t}{}_{\bar r}{}_{u}{}_{\bar s} 
\nn \\ &
+ 48(1 + 
a_1)\o^\tbzero_i{}^{\bar 
m}{}^{n}\o^\tbzero_j{}_{n}{}^{r}R^\tbzero{}^{\bar r}{}_{\bar m}{}_{r}{}^{s}R^\tbzero{}^{ \bar 
s}{}^{t}{}_{s}{}^{u}R^\tbzero{}_{t}{}_{\bar r}{}_{u}{}_{\bar s} 
+ 48(1 + 
a_2)\o^\tbzero_i{}^{\bar 
m}{}^{n}\o^\tbzero_j{}^{\bar r}{}^{r}R^\tbzero{}^{ \bar s}{}_{\bar 
m}{}_{s}{}^{t}R^\tbzero{}_{n}{}^{s}{}_{r}{}^{u}R^\tbzero{}_{t}{}_{\bar r}{}_{u}{}_{\bar 
s} 
\nn \\ &
+ 96(1 + \
a_2)\o^\tbzero_i{}^{\bar 
m}{}^{n}\o^\tbzero_j{}^{\bar r}{}^{r}R^\tbzero{}^{ \bar s}{}_{\bar 
m}{}_{n}{}^{s}R^\tbzero{}_{r}{}^{t}{}_{s}{}^{u}R^\tbzero{}_{t}{}_{\bar r}{}_{u}{}_{\bar 
s} 
- 48(1 + 
a_1)\o^\tbzero_i{}^{\bar 
m}{}^{n}\o^\tbzero_j{}_{n}{}^{r}R^\tbzero{}^{\bar 
r}{}^{s}{}_{r}{}^{t}R^\tbzero{}_{s}{}^{u}{}_{t}{}^{v}R^\tbzero{}_{u}{}_{\bar 
m}{}_{v}{}_{\bar r} 
\nn \\ &
+ 48(1 + 
a_2)\o^\tbzero_i{}^{\bar 
m}{}^{n}\o^\tbzero_j{}^{\bar 
r}{}^{r}R^\tbzero{}_{n}{}^{s}{}_{r}{}^{t}R^\tbzero{}_{s}{}^{u}{}_{t}{}^{v}R^\tbzero{}_{u}{}_{\bar 
m}{}_{v}{}_{\bar r} \brs *^\tbzero 1
\ea
Where we see directly that in the reduction  $\cZ_3 = \cZ_4  = \cZ_5 = \cZ_6 = 0$. 
The result above represents the only terms in the reduction result that can not be expressed in terms of $Z_{m \bar m n \bar n}$ for arbitrary choice of the parameters $a_1$ and $a_2$. For this reason we now make the choice $a_1 = a_2$ which then allows the result to be rewritten as  
\ba
\int &\wh s_{18} ( \wh \na \wh  G)^2 \wh R^2  \wh * 1  |_\text{kin} = 
\fr{1}{2 \k_{11}^2} 192 (1 + a_1)  
\int_{\cM_3} F^i \we * F^j \int_{Y_4}  (  i Z_{m \bar n} \o^\tbzero_{i}{}^{\bar r m} \o^\tbzero_{j}{}^{\bar n}{}_{\bar r } + 2  Z_{m \bar n r \bar s} \o_i^\tbzero{}^{\bar n m} \o_j^\tbzero{}^{\bar s r} ) *^\tbzero 1\, . 
\ea

Furthermore we note that if the basis \eqref{def-Bi} is reduced with and arbitrary set of coefficients and then we demand that the result can be expressed in terms of $Z_{m \bar m  n \bar n}$, then only a multiple of the linear combination 
\ba
\label{NablaGContributions}
\int_{\cM_3} F^i \we * F^j \int_{Y_4}  (  i Z_{m \bar n} \o^\tbzero_{i}{}^{\bar r m} \o^\tbzero_{j}{}^{\bar n}{}_{\bar r } + 2  Z_{m \bar n r \bar s} \o_i^\tbzero{}^{\bar n m} \o_j^\tbzero{}^{\bar s r} ) *^\tbzero 1 \,, 
\ea
is produced.

\end{appendix}



\end{document}